\documentclass[11pt]{article}
\usepackage{moriond,epsfig}

\def\be{\begin{equation}}
\def\ee{\end{equation}}
\def\bea{\begin{eqnarray}}
\def\eea{\end{eqnarray}}

\def\cR{{\cal{R}}}
\def\cF{{\cal{F}}}
\def\as{\alpha_{\mbox{\scriptsize s}}}

\begin{document}
\begin{flushright}
{
  IPPP--01--25\\
  DCTP--01--50\\
  hep-ph/0205120 \\
  May 2002
}
\end{flushright}

\vspace*{2cm}
\title{SEMI NUMERICAL RESUMMATION OF EVENT SHAPES AND JET RATES}

\author{GIULIA ZANDERIGHI}

\address{Institute of Particle Physics Phenomenology, University of Durham, Durham DH1 3LE, England}

\maketitle\abstracts{
For many observables, the most difficult part of a single logarithmic
resummation is the analytical treatment of the observable's dependence
on multiple emissions.  We present a general numerical method, which
allows the resummation specifically of these single logarithms. Some
first applications and new results for the thrust major, the
oblateness and the two-jet rate in the Durham algorithm are also
presented.
}
\section{Introduction}
The formation of jets in hadronic collisions is surely one of the most
striking phenomena in high energy collisions.
Their study allows a precise measurement of $\alpha_s$, is important
for any search for new physics and is useful to test our understanding
of strong interaction dynamics.
Generally jet cross sections are quite large but it might be difficult
to get an insight into details of the interaction mechanics. One way
is provided by the study of event shapes and jet rates.
\section{Resummation of event shape distributions}\label{subsec:res}
Event shape variables can be studied at variety of levels, either
through relatively inclusive properties, such as their mean value, or
more exclusively by examining integrated (or differential)
distributions.
The integrated distribution~\footnote{From here onwards, capital
letters denote the observable, while lower case the numerical value.} 
$\Sigma(v)$ accumulates contributions both from real and virtual
emissions.
In spite of the fact that both contributions are separately divergent,
infrared safety of the observable ensures that their sum is finite at
any order in perturbation theory. In the (more inclusive) $v\sim 1$
region the observable is sensitive only to the total energy and
momentum flow, while in the (less inclusive) region $v\to 0$ the
mismatch between real and virtual emissions gets emphasized, so that
divergences do still cancel but large logarithmic enhanced
contributions are produced, which need to be resummed at any order in
perturbation theory.
\subsection{Classification of single logarithmic terms}\label{subsec:slterms}
In spite of the fact that resummation programmes are quite cumbersome
it is possible to classify systematically in a simple way all leading
(DL) and subleading (SL) contributions.
DL contributions are due to emissions which are both soft and
collinear, while SL corrections are found to be due to
\begin{itemize}
\item 
soft, large angle emissions;
\item 
hard, collinear radiation;
\item 
effects due to the running of $\alpha_s$;
\item 
multiple emission effects, i.e. taking into account properly how all
emissions add up together and interfere to contribute to the value of
the observable.
\end{itemize}
As far as the first three items are concerned, to obtain a resummed
prediction at SL level it is enough to understand how a single
emission affects the observable and to exponentiate this result in a
naive way.
Therefore the last point turns out to be the only non trivial task in
any resummation programme and it is exactly this contribution which we
address here.
\subsection{Standard treatment of multiple emission effects}
To understand how multiple emission effects are treated usually let us
consider for instance the single jet broadening $B$ (in the following
$\vec{k}_{ti}$ denote the secondary transverse momenta)
\begin{equation}
2 B Q = \sum_i k_{ti}\>,\quad \mbox{with}\quad k_{ti}=|\vec{k}_{ti}|\>.
\end{equation}
Resummation is achieved factorizing matrix element and the phase space
(using an independent emission
picture~\cite{CTTW,NewBroad})~\footnote{Actually, the single jet broadening,
taken here for illustration, is a non-global observable, so that an
independent emission picture mistreats some of the single
logs~\cite{Nonglobal}.} and the definition of the observable via a
Mellin transform
\begin{equation}
\Theta(2 B Q -\sum_i k_{ti}) 
=\int_{\cal{C}}\frac{d\nu}{2 \pi i \nu} e^{2 \nu B Q} e^{-\nu k_{ti}}\>.
\end{equation}
But, actually, at SL level one needs to take into account the fact
that due to all secondary emissions the hard parton will take a
recoil $p_t$. Therefore the $\Theta$-function needs to be replaced by
\begin{equation}
\Theta(2 B Q -\sum_i k_{ti}) 
\Rightarrow \int d^2 p_t\, \delta^2(\vec{p}_t+\sum_i \vec{k}_{ti})
\,\Theta(2 B Q -\sum_i k_{ti}-p_t)\>. 
\end{equation}
To factorize the $\delta$-function one needs now an additional
two-dimensional Fourier transform. In the case of the thrust
minor~\cite{tmin}, which is defined as the radiation out of a plane,
to fix kinematically the plane one needs an additional
five-dimensional integral, so that generally an analytical treatment
can become very involved.
Furthermore, in certain cases, as for the thrust major, the oblateness
and the 3-jet resolution, an analytical treatment seems to be
unfeasible.
\section{Numerical treatment of multiple emission effects}
Since multiple emission effects have the same origin it is natural to
look for a general approach to the problem.
The idea is to relate multiple emission effects of the observable $V$
under study to a 'simple' reference variable $V_s$ which has the same
double logarithmic structure, i.e. the same dependence on one
soft-collinear emission, but which factorizes in a trivial way.

For an observable given by the sum of $n$ contributions, i.e. sensitive
to all emissions, the natural choice of the simple variable becomes 
\begin{equation}
V=\sum_i v(k_i)\quad \Rightarrow\quad V_s = \max[v(k_1),\dots,v(k_n)]\>. 
\end{equation}
By construction $V_s$ is sensitive only to the 'largest' emission and
this definition also ensures that $V_s$ factorizes in a simple way
(i.e. needs no Mellin transforms)
\begin{equation}
\Theta(V_s- \max[v(k_1),\dots,v(k_n)])= 
\prod_i \Theta(V_s- v(k_i))\>. 
\end{equation}
As a consequence the resummation of $V_s$ becomes straightforward. For
instance for the case of $e^+ e^-\rightarrow 2\, \mbox{jets}$ the
resummed distribution becomes (the case $e^+ e^-\rightarrow 3\,
\mbox{jets}$ is also trivial)
\begin{equation}
\Sigma_s(v_s) = e^{-R_s(v_s)}\,\qquad 
\cR_s(v_s)=C_F\int^{Q^2}\frac{d^2k_t}{\pi k_t^2}
\frac{\as(k_t)}{2\pi}\int_{-\ln 
  \frac{Q e^{-3/4}}{k_t}}^{\ln \frac{Qe^{-3/4}}{k_t}}d\eta\, 
\Theta(v(k)-v_s)\>.
\end{equation}
The aim is now to exploit the 'simple', known distribution $\Sigma_s$
to compute the real observable $\Sigma$.

The two differential distributions $D=d\Sigma/d\ln v$ are related by a
simple convolution
\begin{equation}\label{eq:conv}
\frac{D(v)}{v}=\int\frac{dv_s}{v_s}D_s(v_s)P(v|v_s)\>,
\end{equation}
where $P(v|v_s)$ denotes the conditional probability of having a value
of $v$ for $V$ given a value of $v_s$ for $V_s$.
The two distributions have the same double logarithmic structure
so that $v\sim v_s$ and one can expand $D_s(v_s)\sim
e^{-\cR_s(v_s)}$ around the value of $v_s$
\begin{equation}\label{eq:exp}
D_s(v_s)=_{SL}D_s(v)e^{-R'\ln\frac{v}{v_s}}\>,
\end{equation}
where $R'\equiv - dR/d\ln v$ is a single logarithmic function.
Combing eq. (\ref{eq:conv}) and (\ref{eq:exp}) one obtains
\begin{equation}
D(v)=_{SL} D_s(v)\cF(R')\>,
\end{equation}
with
\be
\cF(R')=\int\frac{dx}{x}e^{-R'\ln x}p(x,R')\>, \qquad x=v/v_s\>, 
\qquad p(x,R')= v\, P(v/v_s)\>.
\ee
At SL level the same relation holds then for the integrated
distribution
\be
\Sigma(v)=\Sigma_s(v)\cF(R')\>.
\ee
This expression shows explicitly that all non trivial multiple
effects can be factored out and embodied in $\cF$. The aim is then to
compute $\cF$ i.e. $p(x,R')$ in a general way.
\subsection{Numerical implementation: the procedure to get $\cF$}
To compute the function $\cF$ numerically one writes a Monte
Carlo~\cite{AGG} which starts from an arbitrary Born configuration
with an arbitrary number of hard emitting legs and
\begin{itemize}
\item[$0$] 
generates the first 'largest' emission with $v(k_1)=v_s$
(i.e. one starts fixing $V_s=v_s$);
\item[$I$] 
generates an additional soft-collinear (SC) emission (according to the
phase space $R'$) which satisfies $v(k_i)< v(k_{i-1})$;
\item[$II$] 
if $v(k_{i})<\epsilon\ll1$, where $\epsilon$ defines the precision of
the procedure, the emission (and all subsequent ones) is too small to
affect the value of the observable, therefore the procedure is stopped
and the momenta generated are passed to a standard routine which
computes the value of $V$; otherwise one sets $i\rightarrow i+1$ and
goes back to step $I$.
\end{itemize}
\section{Recollection of known results and new results}
As a first check of the procedure we compared some Monte Carlo results
with results from the literature
%
for the function $\cF$ for the thrust~\cite{CTTW} and the total and
wide broadening~\cite{NewBroad} and for the thrust minor~\cite{tmin}.
%
Comparisons show that our numerical results turn out to be
interchangeable with analytical ones~\cite{AGG}.
%

We then applied the procedure to obtain some new predictions for 
thrust major and the oblateness, which have so far defied
analytical resummation and to the two-jet rate in the Durham
algorithm, for which only a subset of the single logs had up to now
been calculated~\cite{JetRates,DissSchmell}.

We first used the Monte Carlo procedure to compute the function $\cF$
for these variables, combining this with $\Sigma_s$ we get the first
full resummed prediction at SL level which we then match with second
order exact results.%

\begin{center}
\begin{minipage}{7.0cm}
\epsfig{file=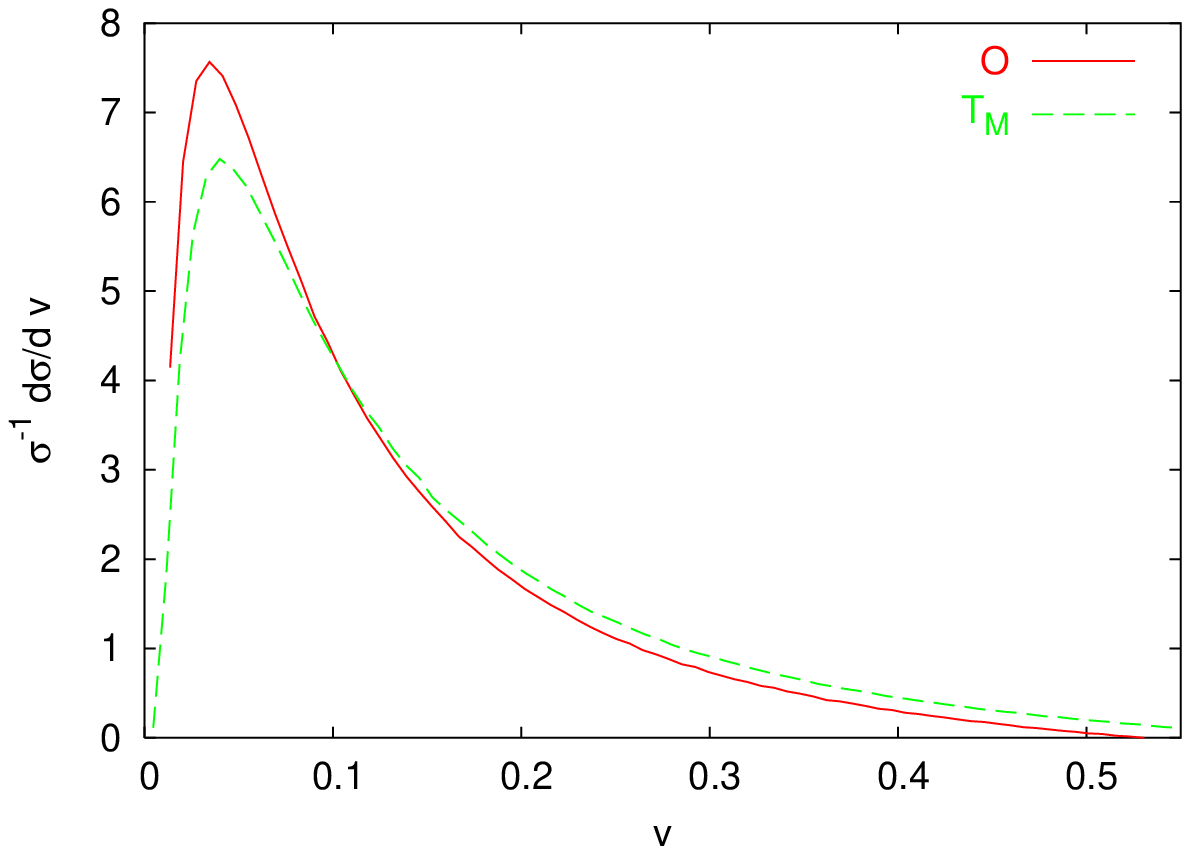,width=1.\textwidth}
\end{minipage}
\begin{minipage}{7.0cm}
\epsfig{file=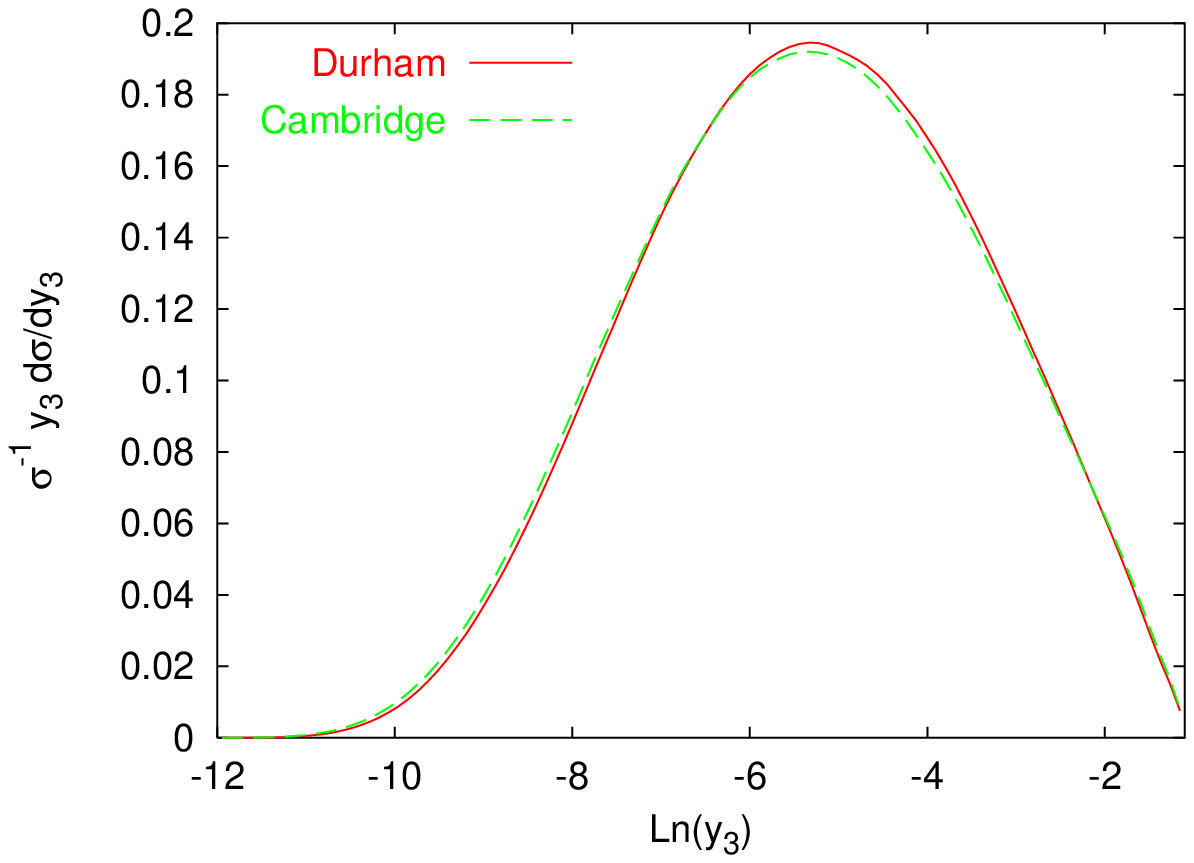,width=1.\textwidth}
\end{minipage}
\end{center}
\section{Conclusions}
The study of event shapes and jet rates has proved to be a powerful
tool to test QCD and to measure $\alpha_s$. Many variables have been
studied in the last years, but never in a general way.  This work is a
first step towards a complete numerical resummation of any $n$-jet
observable in any hard process.
What has been done so far is to implement multiple emission effects
generally, to test the procedure and to obtain some new predictions.
What remains to be done is to make the procedure completely automatic,
i.e. userfriendly and to exploit the method to obtain a bunch of new
predictions especially in hadron-hadron collisions, which will be the
main object of most experimental analyses in the near future.
\section*{Acknowledgments}
It was a pleasure to carry out this work together with Andrea Banfi
and Gavin Salam.
Once more, I'm very grateful to Pino Marchesini and Yuri Dokshitzer
for precious hints and stimulating discussions.
\end{document}